\documentstyle[12pt]{article}

\catcode`@=12
\begin{document}
%----------------------------------------------------------------------
\begin{flushright} UCRHEP-T155\\January 1996\
\end{flushright}
\vspace{0.2in}
\begin{center}
{\Large \bf New Low-Energy Realization of\\The Superstring-Inspired E$_6$ 
Model\\}
\vspace{0.2in}
%----------------------------------------------------------------------
{\bf Ernest Ma \\}
\vspace{0.1in}
{\em Department of Physics, University of California\\}
{\em Riverside, California 92521, USA\\}
\vspace{0.2in}
\end{center}
%----------------------------------------------------------------------
\begin{abstract}
The superstring-inspired E$_6$ model is reduced to the supersymmetric 
standard model extended by a specific U(1) factor.  This choice allows 
for the existence of naturally light singlet neutrinos which also mix 
with $\nu_e$, $\nu_\mu$, and $\nu_\tau$, thus making it possible to 
accommodate all present neutrino data.  Other consequences of this model 
are also discussed: oblique corrections from Z-Z' mixing, phenomenology 
of the two-Higgs-doublet sector, and possible scenarios of gauge-coupling 
unification.
\end{abstract}
%%%%%%%%%%%%%%%%%%%%%%%%%%%%%%%%%%%%%%%%%%%%%%%%%%%%%%%%%%%%%%%%%%%%%%%%
%\psdraft
%%%%%%%%%%%%%%%%%%%%%%%%%%%%%%%%%%%%%%%%%%%%%%%%%%%%%%%%%%%%%%%%%%%%%%%%
%%%%%%%%%%%%%%%%%%%%%%%%%%%%%%%%%%%%%%%%%%%%%%%%%%%%%%%%%%%%%%%%%%%%%%%%
\section{Introduction}\label{sec:intro}
There are a number of possible low-energy realizations of the 
superstring-inspired E$_6$ model\cite{1}.  Their particle content is 
often that given by the fundamental {\bf 27} representation of E$_6$ 
and the gauge group is often larger than that of the standard model by 
at least one U(1) factor.  In particular, there are two neutral singlets 
in each {\bf 27} representation, $N$ and $S$, which may be considered 
as ``neutrinos" because they are very weakly interacting.  The former ($N$) 
is the Dirac partner of an ordinary doublet neutrino $\nu$, whereas the 
latter ($S$) is not.  A possible scenario is that both $N$ and $S$ are heavy 
and that $\nu$ is light via the seesaw mechanism.  In that case, the neutrino 
sector is equivalent to that of the standard model with very small Majorana 
neutrino masses.  On the other hand, there are experimental indications at 
present for three types of neutrino oscillations: solar\cite{2}, 
atmospheric\cite{3}, and laboratory\cite{4}, requiring three mass 
differences which are not possible with only three neutrinos.  Adding a 
fourth doublet neutrino is not allowed because the invisible width of the 
$Z$ boson tells us that there should be only three such neutrinos.  Hence 
the idea of one or more naturally light singlet neutrinos should be 
entertained, and in this talk I will show how a properly chosen extra 
U(1) factor allows for this possibility\cite{5} and discuss also some of the 
other features of this new model.
%%%%%%%%%%%%%%%%%%%%%%%%%%%%%%%%%%%%%%%%%%%%%%%%%%%%%%%%%%%%%%%%%%%%%%%%
\section{Reduction of the Superstring-Inspired E$_6$ Model}\label{sec:reduc}
The fundamental {\bf 27} representation of E$_6$ may be decomposed 
according to its maximum subgroup $SU(3)_C \times SU(3)_L \times SU(3)_R$:
\begin{equation}
{\bf 27} = (3,3,1) + (3^*,1,3^*) + (1,3^*,3).
\end{equation}
Under the reductions $SU(3)_L \rightarrow SU(2)_L \times U(1)_{Y_L}$ and 
$SU(3)_R \rightarrow U(1)_{T_{3R}} \times U(1)_{Y_R}$, the individual 
left-handed fermionic components are defined as follows\cite{6}.
\begin{equation}
(u,d) \sim (3;2,{1 \over 6};0,0), ~~~ u^c \sim (3^*;0,0;-{1 \over 2},
-{1 \over 6}), ~~~ d^c \sim (3^*;0,0;{1 \over 2},-{1 \over 6}),
\end{equation}
\begin{equation}
(\nu_e,e) \sim (1;2,-{1 \over 6};0,-{1 \over 3}), ~~~ e^c \sim (1;0,
{1 \over 3};{1 \over 2},{1 \over 6}), ~~~ N \sim (1;0,{1 \over 3};
-{1 \over 2},{1 \over 6}),
\end{equation}
\begin{equation}
(\nu_E,E) \sim (1;2,-{1 \over 6};-{1 \over 2},{1 \over 6}), ~~~ (E^c,N_E^c) 
\sim (1;2,-{1 \over 6};{1 \over 2},{1 \over 6}),
\end{equation}
\begin{equation}
h \sim (3;0,-{1 \over 3};0,0), ~~~ h^c \sim (3^*;0,0;0.{1 \over 3}), ~~~ 
S \sim (1;0,{1 \over 3};0,-{1 \over 3}).
\end{equation}
The electric charge is given here by
\begin{equation}
Q = T_{3L} + Y_L + T_{3R} + Y_R,
\end{equation}
and three families of the above fermions and their bosonic superpartners 
are assumed.

There are two possible $SO(10)$ subgroups which also contain the $SU(5)$ 
subgroup which contains the standard $SU(3)_C \times SU(2)_L \times U(1)_Y$ 
gauge group.  The conventional decomposition is
\begin{equation}
E_6 \rightarrow SO(10)_A \times U(1)_{\psi_A},
\end{equation}
such that the {\bf 27} splits up as follows:
\begin{eqnarray}
(16,1)_A &=& (u,d) + u^c + e^c + d^c + (\nu_e,e) + N, \\ (10,-2)_A &=& 
h + (E^c,N_E^c) + h^c + (\nu_E,E), \\ (1,4)_A &=& S.
\end{eqnarray}
Note that all the usual quarks and leptons are contained in $(16,1)_A$, 
and the Higgs bosons are in $(10,-2)_A$.  The next step of the 
decomposition is
\begin{equation}
SO(10)_A \rightarrow SU(5) \times U(1)_{\chi_A},
\end{equation}
such that
\begin{equation}
16_A = (10,1) + (5^*,-3) + (1,5), ~~~ 10_A = (5,-2) + (5^*,2), ~~~ 
1_A = (1,0).
\end{equation}
Note that $U(1)_{\psi_A} \times U(1)_{\chi_A}$ breaks down to $U(1)_\eta$ 
via the adjoint {\bf 78}, resulting in $Q_\eta = T_{3L} + 5 Y_L - Q$.

The alternative decomposition\cite{7} is
\begin{equation}
E_6 \rightarrow SO(10)_B \times U(1)_{\psi_B},
\end{equation}
with
\begin{eqnarray}
(16,1)_B &=& (u,d) + u^c + e^c + h^c + (\nu_E,E) + S, \\ (10,-2)_B &=& 
h + (E^c,N_E^c) + d^c + (\nu_e,e), \\ (1,4)_B &=& N.
\end{eqnarray}
The next step is then
\begin{equation}
SO(10)_B \rightarrow SU(5) \times U(1)_{\chi_B},
\end{equation}
such that
\begin{equation}
16_B = (10,1) + (5^*,-3) + (1,5), ~~~ 10_B = (5,-2) + (5^*,2), ~~~ 
1_B = (1,0).
\end{equation}
Note that $S$ is trivial under $U(1)_{\chi_A}$, whereas $N$ is trivial 
under $U(1)_{\chi_B}$.

In the E$_6$ superstring-inspired model, the Yukawa terms are supposed 
to be restricted to only those contained in {\bf 27}$^3 
\rightarrow 1$, namely $u^c (u N_E^c - d E^c)$, $d^c (u E - d \nu_E)$, 
$e^c (\nu_e E - e \nu_E)$, $S h h^c$, $S (E E^c - \nu_E N_E^c)$, and 
$N (\nu_e N_E^c - e E^c)$.  In the following, I also assume a Z$_2$ 
discrete symmetry where all superfields are odd, except one copy each of 
$(\nu_E,E)$, $(E^c,N_E^c)$, and $S$, which are even.  The bosonic 
components of the even superfields will serve as Higgs bosons which 
break the electroweak gauge symmetry.  Specifically, $\langle \tilde N_E^c 
\rangle$ generates $m_u$, $m_D$, and $m_1$; $\langle \tilde \nu_E \rangle$ 
generates $m_d$, $m_e$, and $m_2$; and $\langle \tilde S \rangle$ 
generates $m_h$ and $m_E$.  The mass matrix spanning $\nu_e$, $N$, 
$\nu_E$, $N_E^c$, and $S$ is then given by
\begin{equation}
{\cal M} = \left[ \begin{array} {c@{\quad}c@{\quad}c@{\quad}c@{\quad}c} 
0 & m_D & 0 & 0 & 0 \\ m_D & 0 & 0 & 0 & 0 \\ 0 & 0 & 0 & m_E & m_1 \\ 
0 & 0 & m_E & 0 & m_2 \\ 0 & 0 & m_1 & m_2 & 0 \end{array} \right].
\end{equation}
%%%%%%%%%%%%%%%%%%%%%%%%%%%%%%%%%%%%%%%%%%%%%%%%%%%%%%%%%%%%%%%%%%%%%%%%
%%%%%%%%%%%%%%%%%%%%%%%%%%%%%%%%%%%%%%%%%%%%%%%%%%%%%%%%%%%%%%%%%%%%%%%%
\section{Naturally Light Singlet Neutrinos}\label{sec:light}
If the E$_6$ superstring breaks only via the flux mechanism, then it is 
not possible to have only the standard-model gauge group.  The latter must be 
extended by at least the $U(1)_\eta$ factor mentioned previously.  Under 
$U(1)_\eta$, $N$ and $S$ transform identically [and so do $(\nu_e,e)$ and 
$(\nu_E,E)$ as well as $d^c$ and $h^c$] but nontrivially, hence they are 
forbidden to have Majorana mass terms $N^2$ and $S^2$.  Therefore, in the 
above mass matrix $\cal M$, $\nu_e$ and $N$ together make up a Dirac 
neutrino of mass $m_D$, whereas $m_S \sim 2 m_1 m_2 / m_E$.  If $U(1)_\eta$ 
is actually broken at some large scale, then $N$ and $S$ may acquire large 
Majorana masses through gravitationally induced nonrenormalizable 
interactions\cite{8}.  In that case, $N$ and $S$ are superheavy, whereas 
$m_\nu \sim m_D^2 / m_N$.

To obtain naturally light singlet neutrinos, it is now obvious that 
$U(1)_\eta$ should be replaced with $U(1)_N (\equiv U(1)_{\chi_B})$ 
under which
\begin{equation}
N \sim 0, ~~~ S \sim 5, ~~~ (u,d),~u^c,~e^c \sim 1, ~~~ d^c,~(\nu_e,e) \sim 2, ~
~~ h,~(E^c,N_E^c) \sim -2, ~~~ h^c,~(\nu_E,E) \sim -3.
\end{equation}
Now it is possible to have both light doublet neutrinos: $m_\nu \sim 
m_D^2/m_N$, and light singlet neutrinos: $m_S \sim 2 m_1 m_2/m_E$. 
In addition, the $\nu_e N_E^c - e E^c$ mass term allows the two types 
of neutrinos to mix.  Note that
\begin{equation}
Q_N = 6 Y_L + T_{3R} - 9 Y_R.
\end{equation}
%%%%%%%%%%%%%%%%%%%%%%%%%%%%%%%%%%%%%%%%%%%%%%%%%%%%%%%%%%%%%%%%%%%%%%%%
\section{Oblique Corrections from Z-Z' Mixing}\label{sec:obliq}
The proposed symmetry breaking is achieved through a combination of the 
adjoint {\bf 78} which breaks E$_6$ down to $SU(3)_C \times SU(2)_L \times 
U(1)_{Y_L} \times U(1)_{T_{3R}} \times U(1)_{Y_R}$ and a pair of 
superheavy {\bf 27} and {\bf 27}$^*$ which break it down to $SO(10)_B$, 
resulting in the intermediate gauge group
\begin{equation}
G = SU(3)_C \times SU(2)_L \times U(1)_Y \times U(1)_N.
\end{equation}
Then $U(1)_N$ is broken by $\langle \tilde S \rangle \equiv u$, and 
$SU(2)_L \times U(1)_Y$ breaks down to $U(1)_Q$ via $\langle \tilde \nu_E 
\rangle \equiv v_1$, and $\langle \tilde N_E^c \rangle \equiv v_2$. 
Let
\begin{equation}
v^2 = v_1^2 + v_2^2, ~~~ \tan \beta = {v_2 \over v_1}, ~~~ g_Z^2 = g_1^2 
+ g_2^2,
\end{equation}
then the observed $Z$ boson is a linear combination of $Z_1$ from 
$SU(2)_L \times U(1)_Y$ and $Z_2$ from $U(1)_N$: $Z = Z_1 \cos \theta 
+ Z_2 \sin \theta$, where
\begin{equation}
M_Z^2 \simeq {1 \over 2} g_Z^2 v^2 \left[ 1 - \left( \sin^2 \beta - 
{3 \over 5} \right)^2 {v^2 \over u^2} \right], ~~~ \theta \simeq 
-\sqrt {2 \over 5}~{g_Z \over g_N} \left( \sin^2 \beta - {3 \over 5} 
\right) {v^2 \over u^2}.
\end{equation}
The deviations from the standard model may be expressed in terms of the 
conventional oblique parameters:
\begin{eqnarray}
\epsilon_1 &=& \alpha T ~=~ \left(\sin^4 \beta - {9 \over 25} \right) {v^2 
\over u^2}, \\ \epsilon_2 &=& - {{\alpha U} \over {4 \sin^2 \theta_W}} ~=~ 
\left( \sin^2 \beta - {3 \over 5} \right) {v^2 \over u^2}, \\ \epsilon_3 
&=& {{\alpha S} \over {4 \sin^2 \theta_W}} ~=~ {2 \over 5} \left( 1 + 
{1 \over {4 \sin^2 \theta_W}} \right) \left( \sin^2 \beta - {3 \over 5} 
\right) {v^2 \over u^2}.
\end{eqnarray}
The present precision data from LEP at CERN have errors of order a few 
$\times 10^{-3}$.  This means that $u \sim$ TeV is allowed.
\section{Phenomenology of the Two-Higgs-Doublet Sector}\label{sec:higgs}
At the electroweak energy scale, there are only two Higgs doublets in this 
model, but they differ\cite{9} from the ones of the minimal supersymmetric 
standard model (MSSM).  The reason is that the superpotential has the 
term $f (\nu_E N_E^c - E E^c) S$ which has no analog in the MSSM.  Let
\begin{equation}
\tilde \Phi_1 \equiv \left( \begin{array} {c} \bar \phi_1^0 \\ -\phi_1^- 
\end{array} \right) \equiv \left( \begin{array} {c} \tilde \nu_E \\ 
\tilde E \end{array} \right), ~~~ \Phi_2 \equiv \left( \begin{array} {c} 
\phi_2^+ \\ \phi_2^0 \end{array} \right) \equiv \left( \begin{array} {c} 
\tilde E^c \\ \tilde N_E^c \end{array} \right), ~~~ \chi \equiv \tilde S,
\end{equation}
then the quartic terms of the Higgs potential are given by the sum of
\begin{equation}
V_F = f^2 [(\Phi_1^\dagger \Phi_2)(\Phi_2^\dagger \Phi_1) + (\Phi_1^\dagger 
\Phi_1 + \Phi_2^\dagger \Phi_2)(\bar \chi \chi)],
\end{equation}
and
\begin{eqnarray}
V_D &=& {1 \over 8} g_2^2 [(\Phi_1^\dagger \Phi_1)^2 + (\Phi_2^\dagger \Phi_2)^2 
+ 2 (\Phi_1^\dagger \Phi_1)(\Phi_2^\dagger \Phi_2) - 4 (\Phi_1^\dagger \Phi_2)
(\Phi_2^\dagger \Phi_1)] \nonumber \\ &+& {1 \over 8} g_1^2 [(\Phi_1^\dagger 
\Phi_1)^2 + (\Phi_2^\dagger \Phi_2)^2 - 2 (\Phi_1^\dagger \Phi_1)
(\Phi_2^\dagger \Phi_2)] \nonumber \\ &+& {1 \over 80} g_N^2 [9 
(\Phi_1^\dagger \Phi_1)^2 + 4 (\Phi_2^\dagger \Phi_2)^2 + 12 (\Phi_1^\dagger 
\Phi_1)(\Phi_2^\dagger \Phi_2) - 30 (\Phi_1^\dagger \Phi_1)(\bar \chi \chi) 
\nonumber \\ &~& ~~~~~~~~ - 20 (\Phi_2^\dagger \Phi_2)(\bar \chi \chi) 
+ 25 (\bar \chi \chi)^2].
\end{eqnarray}
Let $\langle \chi \rangle = u$, then $\sqrt 2 Re \chi$ is a physical 
scalar boson with $m^2 = (5/4) g_N^2 u^2$, and couples to $\Phi_1^\dagger 
\Phi_1$ with strength $\sqrt 2 u (f^2 - (3/8)g_N^2)$.  The effective 
$(\Phi_1^\dagger \Phi_1)^2$ coupling $\lambda_1$ is then given by
\begin{equation}
\lambda_1 = {1 \over 4} (g_1^2 + g_2^2) + {9 \over 40} g_N^2 - {{2 (f^2 - 
(3/8)g_N^2)^2} \over {(5/4) g_N^2}}.
\end{equation}
The other quartic self-couplings of the reduced Higgs potential involving 
only $\Phi_1$ and $\Phi_2$ have similar additional contributions. 
Consequently,
\begin{eqnarray}
\lambda_1 &=& {1 \over 4} (g_1^2 + g_2^2) + {6 \over 5} f^2 - {{8 f^4} \over 
{5 g_N^2}}, \\ \lambda_2 &=& {1 \over 4} (g_1^2 + g_2^2) + {4 \over 5} f^2 
- {{8 f^4} \over {5 g_N^2}}, \\ \lambda_3 &=& -{1 \over 4} g_1^2 + {1 \over 4} 
g_2^2 + f^2 - {{8 f^4} \over {5 g_N^2}}, \\ \lambda_4 &=& -{1 \over 2} g_2^2 
+ f^2.
\end{eqnarray}
The MSSM is recovered in the limit of $f=0$ as expected.  Otherwise, the 
two-Higgs-doublet structure is different.  In particular,
\begin{equation}
(m_h^2)_{\rm max} = 2 v^2 \left[ {1 \over 4} g_Z^2 \cos^2 2 \beta + f^2 
\left( {3 \over 2} + {1 \over 5} \cos 2 \beta - {1 \over 2} \cos^2 2 \beta 
\right) - {{8 f^4} \over {5 g_N^2}} \right] + \epsilon,
\end{equation}
where $\epsilon$ comes from radiative corrections, the largest contribution 
being that of the top quark:
\begin{equation}
\epsilon \simeq {{3 g_2^2 m_t^4} \over {8 \pi^2 M_W^2}} \ln \left( 1 + 
{{\tilde m^2} \over m_t^2} \right).
\end{equation}
Normalizing $g_N^2$ to be equal to $(5/3) g_1^2$ and varying $f^2$,
\begin{equation}
(m_h^2)_{\rm max} < M_Z^2 \left[ \cos^2 2 \beta + {25 \over 24} \sin^2 
\theta_W \left( {3 \over 2} + {1 \over 5} \cos 2 \beta - {1 \over 2} 
\cos^2 2 \beta \right)^2 \right] + \epsilon.
\end{equation}
Numerically, the maximum value occurs at $\beta = 0$, which corresponds 
to $m_h \simeq 140$ GeV, as compared to 128 GeV in the MSSM.
\section{Possible Scenarios of Gauge-Coupling Unification}\label{sec:unifi}
Consider now the issue of gauge-coupling unification.  The evolution 
equations of $\alpha_i \equiv g_i^2 / 4 \pi$ are generically given to 
two-loop order by
\begin{equation}
\mu {{\partial \alpha_i} \over {\partial \mu}} = {1 \over {2 \pi}} \left[ 
b_i + {b_{ij} \over {4 \pi}} \alpha_j (\mu) \right] \alpha_i^2 (\mu),
\end{equation}
where $\mu$ is the running energy scale and the coefficients $b_i$ and 
$b_{ij}$ are determined by the particle content of the model.  To one loop, 
the above equation is easily solved:
\begin{equation}
\alpha_i^{-1} (M_1) = \alpha_i^{-1} (M_2) - {b_i \over {2 \pi}} \ln {M_1 
\over M_2}.
\end{equation}
Below $M_{SUSY}$, assume the standard model with 2 Higgs doublets, then
\begin{equation}
b_1 = {21 \over 5}, ~~~ b_2 = -3, ~~~ b_3 = -7.
\end{equation}
Above $M_{SUSY}$ in the MSSM,
\begin{equation}
b_1 = 3(2) + {3 \over 5} (4) \left( {1 \over 4} \right), ~~~ b_2 = -6 + 3(2) 
+ 2 \left( {1 \over 2} \right), ~~~ b_3 = -9 + 3(2).
\end{equation}
It is now well-known that for $M_{SUSY} \sim 10^4$ GeV, the gauge 
couplings do in fact unify at $M_U \sim 10^{16}$ GeV.

In the present model with $u \sim M_{SUSY}$, 
\begin{equation}
b_1 = 3(3) ~+~ ?, ~~~ b_2 = -6 + 3(3) ~+~ ?, ~~~ b_3 = -9 + 3(3) ~+~ ?, ~~~ 
b_N = 3(3) ~+~ ?
\end{equation}
There are two possible scenarios for gauge-coupling unification.  The first 
is an analog of the MSSM.  Add one extra copy of $(\nu_e,e)$ and $(E^c,N_E^c)$ 
[having $\sum Q_N = 0$ so that the theory remains anomaly-free], then
\begin{equation}
\Delta b_1 = {3 \over 5}, ~~~ \Delta b_2 = 1, ~~~ \Delta b_3 = 0, ~~~ 
\Delta b_N = {2 \over 5}.
\end{equation}
This again leads to $M_U \sim 10^{16}$ GeV, which is actually not so good 
because the string scale is an order of magnitude higher.  Also, it is hard to 
understand theoretically why the chosen superfields are light but their 
companions in the same E$_6$ multiplet are superheavy.  On the other hand, 
this is no worse than the usual assumptions taken in SUSY $SU(5)$ or 
SUSY $SO(10)$.

The second scenario is to insist on having $M_U \sim 5 \times 10^{17}$ GeV, 
and allow some components of the superheavy {\bf 27} and 
{\bf 27}$^*$ multiplets to be somewhat lighter than the others.  In 
particular, take 3 copies of $(u,d) + (u^*,d^*)$ and $(\nu_e,e) + 
(\nu_e^*,e^*)$ with $M'$ much below $M_U$, then between $M'$ and $M_U$,
\begin{equation}
\Delta b_1 = 3 \times \left( {1 \over 5} + {3 \over 5} \right) = {12 \over 5}, 
~~~ \Delta b_2 = 3 \times (3 + 1) = 12, ~~~ \Delta b_3 = 3 \times (2 + 0) = 6, 
\end{equation}
\begin{equation}
\Delta b_N = 3 \times \left( {3 \over 10} + {2 \over 5} \right) = {21 
\over 10}.
\end{equation}
As a result, gauge-coupling unification at the string scale is achieved 
with an intermediate scale of $M' \sim 10^{16}$ GeV.
\section{Conclusions}\label{sec:concl}
In conclusion, the superstring-inspired E$_6$ model provides a framework 
for accommodating naturally light singlet neutrinos as well as naturally 
light doublet neutrinos which also mix with each other.  The key is the 
reduction
\begin{equation}
E_6 \rightarrow SU(3)_C \times SU(2)_L \times U(1)_Y \times U(1)_N,
\end{equation}
under which $N$ is trivial, but $S$ is not.  Hence $N$ may acquire a large 
Majorana mass $m_N$, and the mass matrix $\cal M$ of Eq.~(19) becomes
\begin{equation}
{\cal M} = \left[ \begin{array} {c@{\quad}c@{\quad}c@{\quad}c@{\quad}c} 
0 & m_D & 0 & m_3 & 0 \\ m_D & m_N & 0 & 0 & 0 \\ 0 & 0 & 0 & m_E & m_1 \\ 
m_3 & 0 & m_E & 0 & m_2 \\ 0 & 0 & m_1 & m_2 & 0 \end{array} \right].
\end{equation}
This means that the doublet neutrinos obtain very small masses from the 
usual see-saw mechanism: $m_\nu \sim m_D^2/m_N$, whereas the singlet 
neutrinos $S$ get theirs from an analogous $3 \times 3$ mass matrix: 
$m_S \sim 2 m_1 m_2/m_E$.

Other properties of this model include: (1) the two-Higgs-doublet structure 
at the electroweak energy scale is not that of the minimal supersymmetric 
standard model; (2) an additional neutral gauge boson (Z') is possible at 
the TeV energy scale; and (3) gauge couplings may unify at the string 
compactification scale if there can be variations of masses within some 
superheavy supermultiplets.
\section*{Acknowledgments}
I thank Profs.~G.~Zoupanos and N.~Tracas, and all the other organizers of 
the 5th Hellenic School and Workshops on Elementary Particle Physics for 
their great hospitality and a very stimulating program.  This work was 
supported in part by the U.~S.~Department of Energy under Grant 
No.~DE-FG03-94ER40837.
%%%%%%%%%%%%%%%%%%%%%%%%%%%%%%%%%%%%%%%%%%%%%%%%%%%%%%%%%%%%%%%%%%%%%%%%

%%%%%%%%%%%%%%%%%%%%%%%%%%%%%%%%%%%%%%%%%%%%%%%%%%%%%%%%%%%%%%%%%%%%%%%%
\end{document}